\documentclass{article}
\usepackage{graphicx}
\title{\bf Signature change from Schutz's canonical quantum cosmology and its classical analogue}
\author{Pouria Pedram\thanks{Email: pouria.pedram@gmail.com}, Shahram Jalalzadeh\thanks{Email: s-jalalzadeh@sbu.ac.ir}
\\ {\small Department of Physics, Shahid Beheshti University,
Evin, Tehran 19839, Iran}}
\usepackage{amsmath}

\begin{document}

\maketitle \baselineskip 24pt
\begin{abstract}
We study the signature change in a perfect fluid
Friedmann-Robertson-Walker quantum cosmological model. In this work
the Schutz's variational formalism is applied to recover the notion
of time. This gives rise to a Schr\"odinger-Wheeler-DeWitt equation
with arbitrary ordering for the scale factor. We use the
eigenfunctions in order to construct wave packets and evaluate the
time-dependent expectation value of the scale factor which coincides
with the ontological interpretation. We show that these solutions
exhibit signature transitions from a finite Euclidean to a
Lorentzian domain. Moreover, such models are equivalent to a
classical system where, besides the perfect fluid, a repulsive fluid
is present.
\end{abstract}

\textit{Pacs}:{ 98.80.Qc, 04.40.Nr, 04.60.Ds}

\section{Introduction}
The notion of signature transition mainly started to appear in the
works of Hartle and Hawking
\cite{HartleandHawking1983,Hawking1984,HalliwellandHartle1990,GibbonsandHartle1990},
where they argued that in quantum cosmology, amplitudes for gravity
should be expressed as the sum of all compact Riemannian manifolds
whose boundaries are located at the signature changing hypersurface.
A signature changing spacetime is a manifold which contains both
Euclidean and Lorentzian regions  \cite{Dray,Hartley}. In classical
general relativity, the metric which represents signature change
must be either degenerate (vanishing determinant) or discontinuous.
On the other hand, Einstein's equations implicitly assume that the
metric is non-degenerate and at least continuous \cite{Campbell}.

In more recent times, a number of authors have studied this problem
when a scalar field is coupled to Einstein's field equations and
shown that the resulting solutions, when properly parameterized,
exhibit signature transition \cite{Ellis1,Ellis2,Hayward,Martin}. In
a similar way, a classical model is studied in Ref.~\cite{signature}
in which a self-interacting scalar field is coupled to Einstein's
equations with a Sinh-Gordon interaction potential. The field
equations are solved exactly for the scale factor and scalar field
which give rise to a spatially flat Friedmann-Robertson-Walker (FRW)
cosmology with signature changing properties. The case of the
non-flat Universe is addressed in \cite{Ghafoori} with a discussion
about the conditions under which signature transition exists. It is
well known that in classical signature change spacetimes, we have
some junction conditions on the signature changing hypersurface. On
the other hand, there is no any satisfactory unique junction
condition (see \cite{Mansouri,Ellis3} and references therein). At
the quantum cosmology level the same issue is investigated in
Ref.~\cite{Dereli} with an analysis pertaining to the exact
solutions of the Wheeler-DeWitt equation. Signature transition has
also been studied in multi-dimensional classical and quantum
cosmology \cite{Jalalzadeh} where a $4+d$-dimensional spacetime is
minimally coupled to a scalar field. The coupling between spinor
field and gravity based on FRW and Bianchi cosmological models are
also addressed in Refs.~\cite{jalalzadeh1,jalalzadeh2},
respectively. Also, it has been used as a compactification mechanism
for Kaluza-Klein cosmology \cite{Embacher,Darabi} with a
cosmological constant. Finally, the issue of the classically
signature change in the brane world models is studied in
\cite{brane}.

In this paper, we consider a smooth signature changing type
spacetime. First, we study the classical solutions with one
component perfect fluid and show that the solutions do not extend to
the Euclidian region. Then, we construct the corresponding quantum
cosmological model via Schutz's variational formalism which has been
attracted much attentions in recent times \cite{pedram,monerat}. By
canonical quantization of this model, we can avoid the singularity
and consequently, we obtain the signature changing type metric. At
last, we show that the quantum signature change scenario can be
reproduced exactly by a classical model where a repulsive fluid is
added to the normal perfect fluid. The repulsive fluid can be given
by a stiff matter equation of state $p=\rho$, independently of the
content of the normal fluid.

The paper is organized as follows: In Sec.~\ref{sec2}, the quantum
cosmological model with a perfect fluid as the matter content is
constructed in Schutz's formalism \cite{11}, and the
Schr\"odinger-Wheeler-DeWitt (SWD) equation in minisuperspace is
written down to quantize the model. The wave function depends on the
scale factor $a$ and the canonical variable associated to the fluid,
which in the Schutz's variational formalism plays the role of time
$T$. We separate the wave function into two parts, one depending
solely on the scale factor and the other depending only on the time.
The solution in the time sector of the SWD equation is trivial,
leading to imaginary exponentials of type $e^{iET}$, where $E$ is
the energy of the system. Then, we construct the wave packets from
the eigenfunctions and compute the time-dependent expectation value
of the scale factors. We analyze the results in the context of
signature change and construct the classical analogue of quantum
solutions via adding a repulsive perfect fluid. In Sec.~\ref{sec3},
we present our conclusions.

\section{The model}\label{sec2}
The action for gravity plus perfect fluid in Schutz's formalism is
written as
\begin{eqnarray}
\label{action} \mathcal{A} &=& \int_Md^4x\sqrt{-g}\, R +
2\int_{\partial M}d^3x\sqrt{h}\, h_{ab}\, K^{ab}+
\int_Md^4x\sqrt{-g}\, p \quad ,
\end{eqnarray}
where $h_{ab}$ is the induced metric over the three-dimensional
spatial hypersurface, which is the boundary $\partial M$ of the four
dimensional manifold $M$ and $K^{ab}$ is the extrinsic curvature. We
choose units such that the factor $16\pi G$ becomes equal to one.
The first two terms were first obtained in \cite{7} and the last
term (\ref{action}) represents the matter contribution to the total
action. Perfect fluid satisfies the barotropic equation of state
\begin{equation}
p = \alpha\rho.
\end{equation}
In Schutz's formalism \cite{11} the fluid's four-velocity is
expressed in terms of five potentials $\epsilon$, $\zeta$, $\beta$,
$\theta$ and $S$:
\begin{equation}
u_\nu = \frac{1}{\mu}(\epsilon_{,\nu} + \zeta\beta_{,\nu} + \theta
S_{,\nu}),
\end{equation}
where $\mu$ is the specific enthalpy and $S$ is the specific
entropy. The potentials $\zeta$ and $\beta$ are connected with
rotation and are absent in FRW models and the variables $\epsilon$
and $\theta$ have no clear physical meaning. The four-velocity
satisfies the following normalization condition
\begin{equation}
u^\nu u_\nu = -1.
\end{equation}
The Hartle-Hawking no-boundary proposal
\cite{HartleandHawking1983,HalliwellandHartle1990,GibbonsandHartle1990}
implies that spacetime is partly Euclidian and partly Lorentzian.
The motivation of this proposal involves the path integral
formulation of quantum gravity. To have a feeling about the quantum
theory it is necessary to have an understanding of the associated
classical theory by constructing the classical spacetime with
signature changing structure. In fact, there are two main proposals
for this purpose. In the first proposal, the metric of spacetime is
everywhere non-degenerate but fails to be continuous at the surface
that divides the Euclidian from the Lorentzian region. On the other
hand, in the second proposition, metric is everywhere smooth but is
degenerate at the surface of signature change
\cite{Ellis1,Hayward,Kossowski2,Tucker}.

Here, we are interested to use the second one. The authors of
Ref.~\cite{Kossowski1} have shown that for smooth signature changing
spacetime there exist coordinates such that
\begin{equation}
ds^2 = - N^2(t)tdt^2 +h_{ij}dx^idx^j.
\end{equation}
For this case, Kossowski and Kriele \cite{KossowskiandKriele} have
shown that the energy-momentum tensor of the matter field becomes
bounded if and only if the signature change hypersurface ($\Sigma$)
is totally geodesic and $\partial_th_{ij}=0$ at $\Sigma$ .

To proceed further, let us consider the signature changing FRW
metric as
\begin{equation}\label{metric}
ds^2 = - N^2(t)tdt^2 + a^2(t)g_{ij}dx^idx^j,
\end{equation}
where $N^2>0$ and is now inserted in action (\ref{action}). In this
expression, $N(t)$ is the lapse function and $g_{ij}$ is the metric
on the constant-curvature spatial section. Using the constraints for
the fluid, and after some thermodynamical considerations and
dropping the surface terms, the final reduced action takes the form
\cite{14}.
\begin{equation}
\mathcal{A}= \int dt\left[-6\frac{\dot a^2a}{t^{1/2}N}+ 6\kappa
Nt^{1/2}a + N^{-1/\alpha} t^{-1/2\alpha}a^3\frac{\alpha}{(\alpha +
1)^{1/\alpha + 1}}(\dot\epsilon + \theta\dot S)^{1/\alpha +
1}\exp\left(-\frac{S}{\alpha}\right) \right],
\end{equation}
where dot denotes the derivation with respect to $t$. The reduced
action may be further simplified using canonical methods \cite{14}
resulting in the super-Hamiltonian
\begin{equation}
{\cal H} = - \frac{p_a^2}{24a} -6\kappa a +\frac{
 p_\epsilon^{\alpha + 1}e^S}{a^{3\alpha}}
\end{equation}
where $p_a= -12{\dot aa}/{t^{1/2}N}$, $p_\epsilon = \rho_0 a^3$ and
$\rho_0=\frac{\displaystyle\mu^{1/\alpha}}{\displaystyle(1+\alpha)^{1/\alpha}}e^{-S/\alpha}$
is the rest mass density of the fluid. The following additional
canonical transformations
\begin{eqnarray}
T &=& -p_Se^{-S}p_\epsilon^{-(\alpha + 1)},  \quad p_T
=p_\epsilon^{\alpha + 1}e^S , \quad\nonumber\\
\bar\epsilon &=& \epsilon - (\alpha + 1)\frac{p_S}{p_\epsilon},
\quad \quad  \bar p_\epsilon = p_\epsilon,
\end{eqnarray}
simplify the super-Hamiltonian to
\begin{equation}
{\cal H} = - \frac{p_a^2}{24a} - 6\kappa a +
\frac{p_T}{a^{3\alpha}}\label{EqHamiltonian}
\end{equation}
where the momentum $p_T$ is the only remaining canonical variable
associated with matter. It appears linearly in the
super-Hamiltonian. The parameter $\kappa$ defines the curvature of
the spatial section, taking the values $0, 1,-1$ for a flat,
positive-curvature or negative-curvature Universe, respectively.

The classical dynamics is governed by the Hamilton equations,
derived from Eq. (\ref{EqHamiltonian}) and Poisson brackets as
\begin{equation}
\left\{
\begin{array}{llllll}
\dot{a} =&\{a,t^{1/2}N{\cal H}\}=-\frac{\displaystyle t^{1/2}Np_{a}}{\displaystyle 12a}\, ,\\
 & \\
\dot{p_{a}} =&\{p_{a},t^{1/2}N{\cal H}\}=- \frac{\displaystyle t^{1/2}Np_a^2}{\displaystyle 24a^2}+6t^{1/2}N\kappa  +3\alpha t^{1/2}N\frac{\displaystyle p_T}{\displaystyle a^{3\alpha+1}}\, ,\\

 & \\

\dot{T} =&\{T,t^{1/2}N{\cal H}\}=t^{1/2}Na^{-3\alpha}\, ,\\
 & \\
\dot{p_{T}} =&\{p_{T},t^{1/2}N{\cal H}\}=0\, .\\
& \\
\end{array}
\right. \label{4}
\end{equation}
Choosing the gauge $N=a^{3\alpha}$, we have $T=\frac{2}{3}
t^{\frac{3}{2}}$ and the following constraint equation ${\cal H} =
0$
\begin{eqnarray}
-\frac{\displaystyle 6\dot{a}^2}{\displaystyle
ta^{3\alpha-1}}-6\kappa a^{3\alpha+1}+p_T=0.
\end{eqnarray}
For the flat case ($\kappa=0$), we have the following solutions for
$\alpha\ne1$
\begin{eqnarray}
a(t)=\left\{\sqrt{\frac{\displaystyle p_T}{\displaystyle
6}}(1-\alpha)\right\}^{\frac{2}{3(1-\alpha)}}\,\,t^{\frac{1}{1-\alpha}},
\end{eqnarray}
and $\alpha=1$ (stiff matter)
\begin{eqnarray}\label{stiffsol}
a(t)=a_0e^{\frac{2}{3}\sqrt{\frac{P_T}{6}}t^{3/2}}.
\end{eqnarray}
Stiff matter is a fluid with pressure equal to the energy density
and speed of sound equal to speed of light. Although at $t=0$
(signature changing hypersurface) this solution is finite and the
energy-momentum tensor is bounded, for $t<0$ the scale factor
becomes complex. Moreover, for $\alpha\ne1$ the scale factor
vanishes and the energy-momentum tensor is not bounded. Therefore,
we can not construct the signature changing spacetime using simple
perfect fluid in the classical domain. In fact, in order to have
this type of spacetime in classical cosmology, we need some kinds of
scalar fields with suitable potential term that can provide the
signature change \cite{signature}.

Imposing the standard quantization conditions on the canonical
momenta and demanding that the super-Hamiltonian operator annihilate
the wave function, we are led to the following SWD equation in the
minisuperspace with general factor ordering ($\hbar=1$)
\begin{equation}
\label{sle} %\frac{\partial^2\Psi}{\partial a^2}
\frac{1}{2}\left(\frac{1}{a^i}\frac{\partial}{\partial
a}\frac{1}{a^j}\frac{\partial}{\partial
a}\frac{1}{a^k}+\frac{1}{a^k}\frac{\partial}{\partial
a}\frac{1}{a^j}\frac{\partial}{\partial
a}\frac{1}{a^i}\right)\psi(a,T) - 144\kappa a\psi(a,T) - i24a^{-
3\alpha}\frac{\partial\psi(a,T)}{\partial T} = 0.
\end{equation}
Where $i+j+k=1$. Equation (\ref{sle}) takes the form of a
Schr\"odinger equation $i\partial\Psi/\partial T = {\hat H} \Psi$.
This operator is formally self-adjoint for any choice of the
ordering parameters $i$, $j$, $k$ with the standard inner product
\begin{equation}
(\Phi,\Psi) = \int_0^\infty a^{-3\alpha} \Phi^*\Psi da.
\end{equation}
Moreover, the wave functions should satisfy the restrictive boundary
conditions of which the simplest ones are
\begin{equation}
\label{boundary} \Psi(0,T) = 0 \quad \mbox{or} \quad
\frac{\partial\Psi (a,T)}{\partial a}\bigg\vert_{a = 0} = 0.
\end{equation}
The SWD equation (\ref{sle}) can  be solved by separation of
variables as
\begin{equation}
\psi(a,t) = e^{iET}\psi(a), \label{11}
\end{equation}
where the $a$ dependent part of the wave function ($\psi(a)$)
satisfies
\begin{equation}
\label{sle2} \frac{1}{2}\left(\frac{1}{a^i}\frac{\partial}{\partial
a}\frac{1}{a^j}\frac{\partial}{\partial
a}\frac{1}{a^k}+\frac{1}{a^k}\frac{\partial}{\partial
a}\frac{1}{a^j}\frac{\partial}{\partial
a}\frac{1}{a^i}\right)\psi(a) - 144\kappa a\psi(a) +24Ea^{-
3\alpha}\psi(a) = 0.
\end{equation}
For flat case ($\kappa=0$), this equation reduces to
\begin{equation}
\label{sle3}\left[\frac{1}{2}\left(\frac{1}{a^i}\frac{\partial}{\partial
a}\frac{1}{a^j}\frac{\partial}{\partial
a}\frac{1}{a^k}+\frac{1}{a^k}\frac{\partial}{\partial
a}\frac{1}{a^j}\frac{\partial}{\partial
a}\frac{1}{a^i}\right)+24Ea^{- 3\alpha}\right]\psi(a) = 0.
\end{equation}
Now, using the relation
\begin{equation}
\left(\frac{1}{a^i}\frac{\partial}{\partial
a}\frac{1}{a^j}\frac{\partial}{\partial
a}\frac{1}{a^k}\right)\psi(a) =
a^{-1}\frac{\partial^2\psi(a)}{\partial
a^2}-(2k+j)a^{-2}\frac{\partial\psi(a)}{\partial
a}+k(k+j+1)a^{-3}\psi(a),
\end{equation}
we can rewrite the equation (\ref{sle3}) as
\begin{equation}
\label{sle4}\psi''-a^{-1}\psi'+\left[\frac{1}{2}\bigl(i(i+j+1)+k(k+j+1)\bigr)a^{-2}+24Ea^{-
3\alpha+1}\right]\psi= 0,
\end{equation}
where a prime denotes the derivation with respect to $a$. Equation
(\ref{sle4}) admits a solution under the form of Bessel functions,
leading to the following final expression for the stationary wave
functions
\begin{equation}
\Psi_E(a,T) = e^{iET}a\,\left[c_1J_{l}\left(\frac{\sqrt{96E}}{3(1 -
\alpha)}a^{\frac{3(1 - \alpha)}{2}}\right) +
c_2Y_{l}\left(\frac{\sqrt{96E}}{3(1 - \alpha)}a^{\frac{3(1 -
\alpha)}{2}}\right)\right],
\end{equation}
where
$l=\frac{2\sqrt{1-\frac{1}{2}\left(i(i+j+1)+k(k+j+1)\right)}}{3(1 -
\alpha)}$. Now, the wave packets can be constructed by superposing
these eigenfunctions with the following structure
\begin{equation}
\Psi(a,T) = \int_0^\infty A(E)\Psi_E(a,T)dE.
\end{equation}
We choose $c_2 = 0$, for satisfying the first boundary condition
(\ref{boundary}). By choosing $A(E)$ as a quasi-gaussian weight
factor and defining $r = \frac{\sqrt{96E}}{3(1-\alpha)}$, an
analytical expression for the wave packet can be found
\begin{equation}
\Psi(a,T) = a\int_0^\infty r^{l + 1}e^{-\gamma r^2 +
i\frac{3}{32}(1-\alpha)^2r^{2} T}J_l(ra^\frac{3(1-\alpha)}{2})dr,
\end{equation}
where $\gamma$ is an arbitrary positive constant. The above integral
is known \cite{Luke}, and the wave packet takes the form
\begin{equation}\label{wave}
\Psi(a,T) =
a^{\frac{3}{2}l(1-\alpha)+1}(2B)^{-l-1}e^{-\frac{a^{3(1-\alpha)}}{4B}},
\end{equation}
where $B = \gamma -i\frac{3}{32}(1-\alpha)^2 T$.  Following the many
worlds interpretation of quantum mechanics \cite{everett}, we may
write the expectation value for the scale factor $a$ as
\begin{equation}
\langle a\rangle(T) = \frac{\int_0^\infty
a^{1-3\alpha}\Psi(a,T)^*\Psi(a,T)da} {\int_0^\infty a^{-3\alpha}
\Psi(a,T)^*\Psi(a,T)da} .
\end{equation}
which yields
\begin{equation}\label{solutions}
\langle
a\rangle(t)=\frac{\Gamma(\frac{l}{2}+1)}{\Gamma(\frac{l}{2}+\frac{2-3\alpha}{3-3\alpha})}
\left[\frac{2}{\gamma}\left(\frac{1}{256}(1-\alpha)^4t^3 +
\gamma^2\right)\right]^\frac{1}{3(1-\alpha)} .
\end{equation}
where we have used $T=\frac{2}{3} t^{\frac{3}{2}}$. These solutions,
asymptotically correspond to the flat classical models for the late
times
\begin{equation}
a(t) \propto t^{1/(1-\alpha)}.
\end{equation}

We can also study the situation from the ontological interpretation
of quantum mechanics \cite{holland,nelson}. In this approach the
wave function can be written as
\begin{equation}
\Psi(a,T) = R\, e^{iS},
\end{equation}
where $R$ and $S$ are real functions. Inserting this expression in
the SWD equation (\ref{sle}), for $\kappa=0$ we have
\begin{eqnarray}
\label{hje}
\frac{1}{a^{3\alpha}}\frac{\partial S}{\partial T} - \frac{1}{24a}\biggr(\frac{\partial S}{\partial a}\biggl)^2 + \, Q &=& 0 ,\\
\frac{\partial R}{\partial T} - \frac{1}{12a^{1 -
3\alpha}}\left(\frac{\partial R}{\partial a}\frac{\partial
S}{\partial a} + R \frac{\partial^2S}{\partial
a^2}-\frac{1}{a}R\frac{\partial S}{\partial a}\right) &=& 0 ,
\end{eqnarray}
where
\begin{eqnarray}
Q =
\frac{1}{24a}\frac{1}{R}\left(\frac{\partial^2R}{\partial
a^2}-\frac{1}{a}\frac{\partial R}{\partial
a}\right)+\frac{i(i+j+1)+k(k+j+1)}{48a^{3}},
\end{eqnarray}
is the quantum potential which modifies the Hamilton-Jacobi equation
(\ref{hje}) and the last term is related to factor ordering. When
the quantum potential is more important than the classical
potential, we can expect a behavior deviating from the classical
one. Note that in the present case, since $\kappa = 0$, the
classical potential is zero. The wave function (\ref{wave}) implies
\begin{eqnarray}
R &=& a^{\frac{3}{2}l(1-\alpha)+1}\left[4\gamma^2 +
\left(\frac{3}{16}\right)^2(1 - \alpha)^4T^2\right]^{-(l+1)/2}\,
\exp\left\{-\frac{\gamma a^{3(1 - \alpha)}}{4\left[\gamma^2 + \left(\frac{3}{32}\right)^2(1 - \alpha)^4T^2\right]}\right\}, \\
S &=& - \frac{3}{128}\frac{(1 - \alpha)^2a^{3(1 - \alpha)}T}{
\left[\gamma^2 + \left(\frac{3}{32}\right)^2(1 -
\alpha)^4T^2\right]}
-\left(l+1\right)\arctan\left[\frac{3}{32}\frac{(1 -
\alpha)^2T}{\gamma}\right].
\end{eqnarray}
The Bohmian trajectories, which determine the behavior of the scale
factor, are given by
\begin{equation}
p_a = \frac{\partial S}{\partial a}.
\end{equation}
Using the above definition, the equation for the Bohmian
trajectories becomes
\begin{equation}
\frac{512}{a}\frac{da}{dT} = 3(1 - \alpha)^3\frac{T}{\left[\gamma^2
+ \left(\frac{3}{32}\right)^2(1 - \alpha)^4T^2\right]}
\end{equation}
which can be integrated to
\begin{equation}
\label{bohmain} a(T) = a_0 \biggr[\gamma^2 +
\biggr(\frac{3}{32}\biggl)^2(1 - \alpha)^4T^2\biggl]^\frac{1}{3(1 -
\alpha)},
\end{equation}
where $a_0$ is an integration constant. This result coincides with
the one which is found by computation of the expectation value of
the scale factor (\ref{solutions}). The quantum potential takes the
form
\begin{eqnarray}
Q(a,T) =  \frac{3}{32a^{3\alpha}}\frac{\gamma(1 - \alpha)^2}{\left[\gamma^2 +
\left(\frac{3}{32}\right)^2(1 - \alpha)^4T^2\right]} \left\{\frac{
\gamma a^{3(1 - \alpha)}}{\left[\gamma^2 +
\left(\frac{3}{32}\right)^2(1 - \alpha)^4T^2\right]}-
(l+1)\right\}\nonumber
\\+\frac{2\left(\frac{9}{4}l^2(1-\alpha)^2-1\right)+i(i+j+1)+k(k+j+1)}{48a^{3}}
.
\end{eqnarray}
Now, using the solution (\ref{bohmain}), we can find the the quantum
potential in terms of the scale factor as
\begin{eqnarray}\nonumber
Q(a) &=&-\frac{1}{48 a^{3}}\left[\frac{9}{2}\gamma
a_0^{3(1-\alpha)} (\alpha-1)\left(\gamma a_0^{3(1-\alpha)} -(l+1)
\right)-2\left(\frac{9}{4}l^2(1-\alpha)^2-1\right)-i(i+j+1)-k(k+j+1)
\right]\\  &:=&-\frac{C}{a^{3}}.
\end{eqnarray}
It is clear that the quantum effects become important near $a=0$
and become negligible for large values of the scale factor. The
avoidance of the singularity is due to the repulsive force $F_a =
-\partial Q(a,T)/\partial a$ extracted from the quantum potential.
%**********************************************************************************
To show this, we can write the super-Hamiltonian for the flat case
in Bohmian picture as
\begin{equation}
{\cal H}_B = - \frac{6a\dot{a}^2}{N^2t}+
\frac{p_T}{a^{3\alpha}}-\frac{C}{a^3}.
\end{equation}
 Now, the zero energy condition yields
\begin{equation}
\frac{6}{N^2t}\left(\frac{\dot
a}{a}\right)^2=\frac{p_T}{a^{3(\alpha+1)}}-\frac{C}{a^6}.
\end{equation}
The sign of the left hand side of the above equation  is negative for
the negative values of $t$ and positive for $t>0$. Consequently the sign
of the right hand side changes  as well. Hence the right hand side
vanishes at $t=0$. Now, since we have solutions both for $t<0$ and $t>0$,
there should therefore exist signature changing hypersurface so
that  $a\propto(C/P_T)^{1/3(1-\alpha)}=a_0$. Also it is easy to see from
equation (42) that
the scale factor is less than $a_0$ for negative values of $t$
and grater than $a_0$ for the positive values of $t$.
Hence, this equation predicts the existence of three regions, namely,
a Lorentzian domain, a signature changing hypersurface and an Euclidean
domain. Therefore, since the discussion above is independent of the choice
of $N$, the quantum signature change behavior is in fact gauge independent.

%*********************************************************************************
The solutions (\ref{solutions}) show a continuous transition
from a finite Euclidean domain to the Lorentzian one. It is easy to
show that
\begin{eqnarray}
\partial_t\langle a\rangle|_{t=0}&=&0,\\
\partial_t\partial_t\langle a\rangle|_{t=0}&=&0,
\end{eqnarray}
which satisfy the Kossowski and Kriele mentioned theorem. Hence, in
general, the quantum model predicts a signature change model when
the singularity is approached. Moreover, the quantum effect leads to
a repulsive force which results in a regular transition from
Euclidean to the Lorentzian region.

The above discussion shows that one of the curious features of quantum cosmology  is the use of Riemannian
signature
spaces to explain the origin of the observable Lorentzian signature Universe.
There are various interpretations of this, the simplest of which is that the signature
of the universe was initially Riemannian and then subsequently changed. It may be argued
that the Lorentzian signature is an independent assumption of relativity rather than a
consequence, with the theory being equally valid for Riemannian signature, and that in
a quantum theory of gravity it would be unnatural to impose signature restrictions on
the metric. The question arises as to whether the qualitative predictions of quantum
cosmology can be obtained from purely classical relativity by relaxing the assumption of Lorentzian signature.
 Also in order to understand the quantum theory it
is necessary to have an understanding of the associated classical theory i.e., the theory of classical spacetimes with signature type change.

Now, we try to construct the classical analogous to the quantum
signature change cosmology. One way of implementing a repulsive
phase in classical cosmology is to consider two fluids, one that
acts attractively, and the other that acts repulsively
\cite{Batista}. It is desirable that the repulsive fluid dominates
for the small values of the scale factor, whereas the attractive
fluid dominates for the large values of the scale factor. For the
flat case, we can obtain the possible model in signature change
coordinate (\ref{metric}) as
\begin{equation}
\label{em}\frac{1}{N^2t}\left(\frac{\dot a}{a}\right)^2 = 8\pi
G\left(\rho_M - \rho_Q\right) = \frac{C_1}{a^m} - \frac{C_2}{a^n},
\end{equation}
where $p_M = \alpha_M\rho_M$, $p_Q = \alpha_Q\rho_Q$, $m = 3(1 +
\alpha_M)$ and $n = 3(1 + \alpha_Q)$. The subscripts $M$ and $Q$
stand for ``normal'' matter component and for ``quantum'' repulsive
component, respectively.

Since the normal matter corresponds to $\alpha_M = -1, 0,
\frac{1}{3}$, and it is also desirable that the repulsive component
dominates at small values of the scale factor, we choose $\alpha_Q
> \frac{1}{3}$. With due attention to (42), we choose
a repulsive stiff matter fluid $\alpha_Q = 1$, which leads to $n =
6$. Then, the solution is
\begin{equation}
\label{ce} \frac{1}{N^2t}\left(\frac{\dot a}{a}\right)^2 =
\frac{C_1}{a^{3(1 + \alpha)}} - \frac{C_2}{a^6}.
\end{equation}
This equation can be solved by reparametrizing the time coordinate
as
\begin{equation}
t^{1/2}dt = dT,
\end{equation}
which results in
\begin{equation}
\biggr(\frac{a'}{a}\biggl)^2 = C_1a^{-3(1 - \alpha)} - C_2a^{-6(1 -
\alpha)},
\end{equation}
where the prime means derivation with respect to $T$. This equation
can be easily solved, leading to the following expression for the
scale factor
\begin{equation}
\label{cs} a(T) = \left(\frac{C_1}{C_2}\right)^\frac{1}{3(1 -
\alpha)}\left[\frac{{C_1}^2C_2}{36(1 - \alpha)^2}T^2 +
1\right]^\frac{1}{3(1 - \alpha)},
\end{equation}
which coincides with the quantum mechanical solution with only the
ordinary perfect fluid (\ref{solutions}). Therefore, the quantum
solutions are equivalent to the classical solutions where gravity is
coupled to the same perfect fluid plus a repulsive fluid with a
stiff matter equation of state $p_Q = \rho_Q$.

The comparison between classical (\ref{cs}) and quantum
(\ref{solutions}) solutions can fix $C_1$ and $C_2$ as
\begin{equation}
C_1 =
\left(\frac{\Gamma(\frac{l}{2}+1)}{\Gamma(\frac{l}{2}+\frac{2-3\alpha}{3-3\alpha})}\right)^{1
- \alpha}\frac{3}{8} 3^{1/3}\frac{(1 -
\alpha)^2}{\gamma^{1/3}},\quad \quad C_2 =
\left(\frac{\Gamma(\frac{l}{2}+1)}{\Gamma(\frac{l}{2}+\frac{2-3\alpha}{3-3\alpha})}\right)^{-2(1
- \alpha)}\frac{3}{4} 3^{1/3}\frac{(1 - \alpha)^2}{\gamma^{4/3}}.
\end{equation}
Now we can check the null energy condition ($\rho + p \geq 0$) in
order to investigate the avoidance of the singularity in comoving
coordinate.

The existence of a repulsive term implies that the energy conditions
are violated as the singularity is approached, leading to its
avoidance. If we define $\rho_{eff}$ and $p_{eff}$ as the sum of the
energy and pressure for both attractive and repulsive fluids and
using the solutions (\ref{cs}), with an unimportant absorbtion of
integration constant in the definition of the time coordinate, we
have
\begin{eqnarray}
\begin{array}{cc}
\rho_{eff} + p_{eff} = \frac{4}{a^{6\alpha}}\biggr( -
\frac{a''}{a} + (1 + 3\alpha)\frac{a'^2}{a^2}\biggl)\\
 = \frac{1}{a^{6\alpha}}\frac{8}{3(1 -
\alpha)^2}\biggr[\frac{(1 + \alpha)T^2 - (1 - \alpha)}{(T^2 +
1)^2}\biggl],
\end{array}
\end{eqnarray}
which is negative for $T < \sqrt{\frac{1 - \alpha}{1 + \alpha}}$.
Therefore, for $\alpha<1$ the null energy condition is violated
around the signature change hypersurfaces.

 Repulsive gravitational effects
in classical general relativity can also be generated by self interacting
scalar fields to which an effective energy density and an effective pressure can be associated such that
\begin{eqnarray}
\begin{cases}p_\phi=\frac{\displaystyle\dot{\phi}^2}{\displaystyle2}-U(\phi),\\
\\
\rho_\phi=\frac{\displaystyle\dot{\phi}^2}{\displaystyle2}+U(\phi).
\end{cases}
\end{eqnarray}
A convenient choice for the potential $U(\phi)$ may lead to the
repulsive effect and consequently we have classical signature
change. According to our discussions of need to specific kind of
matter to have signature changing, we assume that the scalar field
is dominated at early Universe and  we choose a specific form of the
potential with an equation of state $p_\phi=\rho_\phi$ near the
signature changing hypersurface. Now, if  the signature changing
hypersurface is located at $t=0$, then in the vicinity of this point
we have $\alpha_\phi\rightarrow 1$, or in the other words
\begin{eqnarray}
\lim_{\phi\rightarrow 0} U(\phi)=0
\end{eqnarray}
On the other hand, to the reason of change of sign of the pressure
and the energy density of the scalar field from Lorentzian to
Euclidian region, the potential term in the neighborhood of
signature changing hypersurface must be an odd function of $t$. An
example of such kind of self interacting potential, which is
introduced by Dereli and Tucker \cite{signature}, is
\begin{eqnarray}
U(\phi)= \Lambda+a\sinh^2(c\phi)+b\sinh(2c\phi),
\end{eqnarray}
where $a$, $b$ and $c$ are constant parameter. The first two terms
in $U(\phi)$ give rise to a Sinh-Gordon scalar interaction. The
third term breaks the symmetry of the potential under
$\phi\rightarrow-\phi$, and is directly responsible for the
signature changing properties of the solutions.
\begin{figure}
\centerline{\begin{tabular}{ccc}
 \includegraphics[width=8cm]{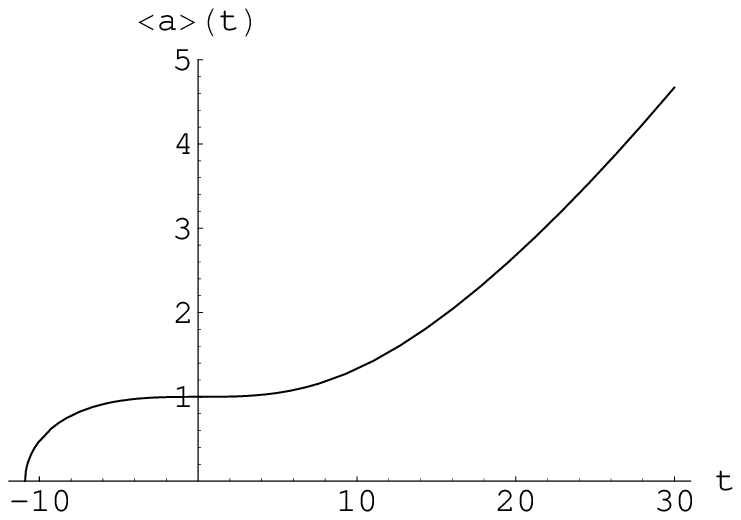}
 &\hspace{2.cm}&
 \includegraphics[width=8cm]{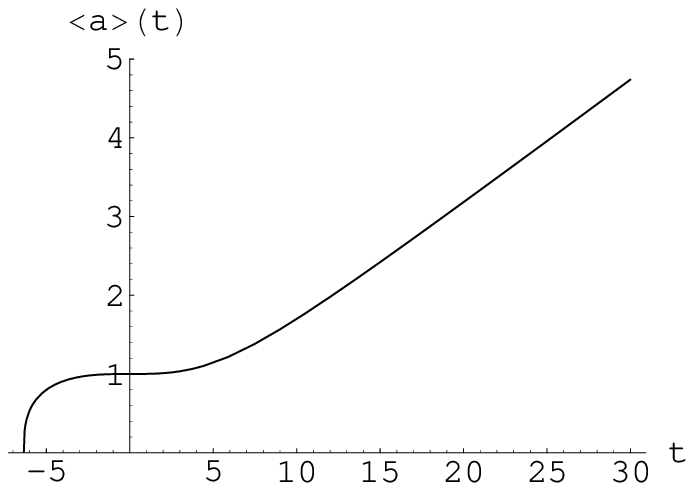}\\
  \includegraphics[width=8cm]{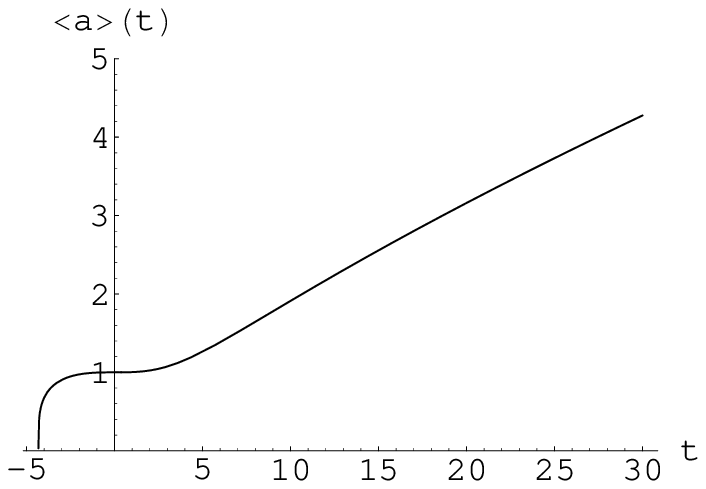}
 &\hspace{2.cm}&
 \includegraphics[width=8cm]{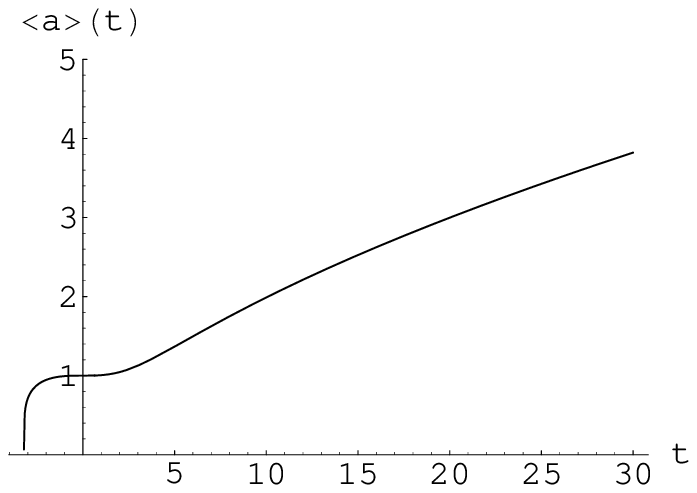}
\end{tabular}  }
\caption{ The expected value of the scale factor for Radiation,
Dust, Cosmic strings and Domain walls dominated Universes (up left
to down right respectively) in $a_0$ unit and $\gamma=1$.}
\label{fig1}
\end{figure}

\section{Conclusions}\label{sec3}
In this work we have investigated a minisuperspace perfect fluid FRW
quantum cosmological model in the context of signature change type
spacetime. The use of Schutz's formalism for perfect fluid allowed
us to obtain a SWD equation in which the only remaining matter
degree of freedom plays the role of time. We found the
eigenfunctions with arbitrary choices of factor ordering. Physically
acceptable wave packets were constructed by appropriate linear
combination of these eigenfunctions. The time evolution of the
expectation value of the scale factor has been determined in the
spirit of the many worlds and ontological interpretations of quantum
cosmology. We have also explored the possibility of having solutions
that are described by degenerate metrics signifying transition from
a Euclidean to a Lorentzian domain at quantum level. Moreover, we
have shown that adding a repulsive stiff matter to the classical
scenario can reproduce the quantum signature changing results.
Finally, we discussed the construction of the self interacting
scalar fields which give rise to classical signature change
scenario.

\end{document}